\documentclass[12pt,preprint]{aastex}

\shorttitle{A New Photo-z Technique}
\shortauthors{Wray \& Gunn}
\begin{document}

\title{A New Technique for Galaxy Photometric Redshifts in the Sloan Digital
       Sky Survey}
\author{James J. Wray\altaffilmark{1}}
\email{jwray@astro.cornell.edu}
\author{James E. Gunn\altaffilmark{2}}
\email{jeg@astro.princeton.edu}

\altaffiltext{1}{Department of Astronomy, Space Sciences Building, Cornell
 University, Ithaca, NY 14853, USA}
\altaffiltext{2}{Princeton University Observatory, Princeton, NJ 08544, USA}

\begin{abstract}
 Traditional photometric redshift methods use only color information about the
 objects in question to estimate their redshifts.  This paper introduces a new
 method utilizing colors, luminosity, surface brightness, and radial light
 profile to measure the redshifts of galaxies in the Sloan Digital Sky Survey
 (SDSS).  We take a statistical approach: distributions of galaxies from the
 SDSS Large-Scale Structure (LSS; spectroscopic) sample are constructed at a
 range of redshifts, and target galaxies are compared to these distributions.
 An adaptive mesh is implemented to increase the percentage of the parameter
 space populated by the LSS galaxies.  We test the method on a subset of
 galaxies from the LSS sample, yielding rms $\Delta z$ of 0.025 for red
 galaxies and 0.030 for blue galaxies (all with $z < 0.25$).  Possible future
 improvements to this promising technique are described, as is our ongoing
 work to extend the method to galaxies at higher redshift.
\end{abstract}

\keywords{galaxies: distances and redshifts --- techniques: photometric ---
 catalogs}

\section{INTRODUCTION}
 Since \citet{hubble29} discovered a linear relationship between the distances
 and redshifts of other galaxies, redshift measurements have been the primary
 method for determining distances to extragalactic objects.  This is 
 normally done using spectra of sufficiently high resolution that individual 
 spectral lines can be resolved and matched to the same features in
 nearby objects, or by matching the spectrum to a model.

 However, measuring the spectrum of an object with  high spectral resolution
 and sufficiently high signal-to-noise
 requires a significantly longer integration time than recording broadband
 photometry of comparable quality.  Thus, it is desirable to be able
 to measure an object's redshift from broadband photometry alone.  Redshifts
 measured this way are called photometric redshifts, or photo-$z$'s.
 Throughout this paper, we will refer to objects for which photo-$z$'s are
 sought as {\em targets}.

 Photo-z techniques date back to \citet{baum62}, who combined nine photometric
 bands to form low-resolution SEDs for elliptical galaxies.  These traced the
 steep 4000 {\AA} break feature, which remains an excellent tool for photo-$z$
 determination since it produces a strong difference in flux between whichever
 two passbands straddle it at a given redshift.  \citet{koo85} was able to
 measure fairly accurate photo-$z$'s for both red and blue galaxies using only
 3 or 4 photometric passbands; his method involved comparisons of observed
 galaxy colors with those predicted by the Bruzual spectral evolution models
 \citep[and companion papers cited therein]{bruzual83a} at a range of
 redshifts.  \citet{connolly95} took a purely empirical (training set-based)
 approach, deriving a correlation between four-band photometric data and the
 measured spectroscopic redshifts of a sample of galaxies.  \citet{sawicki97}
 compare four-band target photometry to that predicted by {\em empirical}
 template spectra.  More recently, {\em hybrid} techniques combining spectral
 template-fitting with training sets have been introduced
 \citep{budavari00, csabai00, csabai03}.

 All of the methods listed above use only the photometric fluxes (i.e. colors
 or apparent magnitudes) of their targets for calculating photo-$z$'s.
 However, galaxy images generally yield additional geometrical information,
 such as angular size, shape, and light distribution (radial and azimuthal).
 In a review of photometric redshift techniques, \citet{koo99} suggested that
 galaxy structural parameters---including surface brightness and radial light
 profile---could be used to reduce the number of passbands needed for precise
 redshift estimates.  Indeed, the bulge-to-total flux ratio was used by
 \citet{sarajedini99} along with $I$-magnitude and $V-I$ color, and
 \citet{kurtz07} have recently developed a novel method that uses only one
 color and the surface brightness from a single band.

 Supervised neural networks have recently been used to compute photo-$z$'s
 from a range of input parameters, including Petrosian radii
 \citep{firth03, vanzella04}, concentration index \citep{collister04},
 surface brightness and axial ratios \citep{ball04}.  \citet{dabrusco07} have
 incorporated Petrosian radii and information about the radial profile into
 their neural network.  \citet{wadadekar05} has
 used a different machine learning method to compute photo-$z$'s based on
 five passband fluxes along with the concentration index, while \citet{way06}
 have used ensemble learning and Gaussian process regression to derive
 photo-$z$'s from colors and various morphological parameters.
 
 This paper introduces a new, statistically-based photo-$z$ technique,
 first conceived by David Schlegel, that uses surface brightness
 and the S\'ersic index---a measure of the radial light profile---in addition
 to five-band photometry.  The method is empirical: the seven properties
 listed are measured for a spectroscopic sample of galaxies, whose redshift
 information is used to estimate photo-$z$'s for the target galaxies.

 Note that photometric redshifts have also been successfully applied to
 quasars \citep[e.g.,][]{richards01, budavari01}.  This paper focuses on
 galaxy photo-$z$'s.

 The paper is structured as follows: in \S2, we describe the spectroscopic
 sample of galaxies used by the photo-$z$ code.  The photo-$z$ technique and
 its development are discussed in \S3, along with other variations that were
 explored.  A test of the photo-$z$ code is described in \S4.  We present our
 conclusions in \S5, and suggest future improvements for increasing the
 accuracy and applicability of the method.

\section{THE SOURCE SAMPLE}
\subsection{SDSS \& the NYU-VAGC}
 As of its Fourth Data Release \citep{adelman06}, the Sloan Digital Sky Survey
 \citep[SDSS;][]{york00,gunn98,gunn06} has imaged roughly 7000 square degrees
 of sky in five bands $(u, g, r, i, z)$ ranging from the near-ultraviolet to
 the near-infrared \citep{fukugita96,smith02}.  Follow-up spectroscopy has been
 performed on objects selected by one of several precisely defined target
 selection algorithms \citep{strauss02, eisenstein01, richards02}.  SDSS has
 measured $\sim10^6$ galaxy spectra, but the number of galaxies detected in
 SDSS imaging is greater by roughly two orders of magnitude.  Thus,
 despite the great size of the SDSS spectroscopic sample, which includes both
 a ``Main'' sample (flux-limited to $r = 17.77$) and a Luminous Red Galaxy
 (LRG) sample (flux- and color-selected, reaching down to $r = 19.5$), the
 huge size of the imaging survey makes it a very attractive target for
 photometric redshift techniques.  Thus we work with SDSS data, although the
 method is in principle applicable to any other imaging survey with similar
 observable parameters.

 The New York University Value-Added Galaxy Catalog
 \citep[NYU-VAGC;][]{blanton05} is essentially an ``extended Main sample;''
 it extends the low-magnitude limit down to $r = 18$, and makes the other
 cuts on the Main sample less restrictive.  It also includes all galaxies
 within 2 arcseconds of any target from the Main, LRG, or QSO samples, and
 thus is useful for analyzing large-scale structure.  In fact, also
 available are subsets of the NYU-VAGC called Large-Scale Structure (LSS)
 samples, which contain only well-characterized galaxies with measured
 spectroscopic redshifts.  These samples are continually updated and
 expanded; we use sample14, which contains
 221,617 galaxies with good photometry.  Specifically, our sample results
 from an apparent magnitude cut, $14.5 < r < 17.5$, an absolute magnitude
 cut, $-23. < M_r < -17.$, and a redshift cut, $0.01 < z < 0.25$.  The
 redshift cut eliminates only a handful of galaxies that are not already
 eliminated by the photometric cuts.

 Finally, the NYU-VAGC also contains a few derived parameters, including
 $K$-corrections and {\em S\'ersic indices} for all galaxies.  The S\'ersic
 index $n$ \citep{sersic68, graham05a} is defined by fitting the radial
 surface brightness profile with a model of the form:
\begin{equation}
 I(r) = A\exp[-(r/r_0)^{1/n}].
\end{equation}
 The value $n=1$ produces an exponential light profile, typical 
 of late-type galaxies (in addition to some low-luminosity early-type
 galaxies), whereas $n=4$ produces a ``de Vaucouleurs profile,'' long
 considered a good description for many early-type
 galaxies.  The SDSS photometric pipeline only performs fits for these
 two particular values, because computing an arbitrary best-fit value
 is computationally very expensive \citep{stoughton02}.
 Thus, \citet{blanton05} calculate this best-fit value of $n$ themselves,
 for each galaxy in the NYU-VAGC 
 (though they do the fits to circularly averaged profiles, whereas the
 SDSS pipeline performs a full 2-dimensional elliptical fit.)

\subsection{Examining the LSS samples}
 \citet{blanton03a} used the slightly older LSS sample12, with cuts very
 similar to the ones we used, to examine correlations among observable
 properties of SDSS galaxies.  The quantities they studied were the four
 colors $u-g, g-r, r-i, i-z$; the absolute magnitude $M_i$; the surface
 brightness $\mu_i$; and the S\'ersic index $n$, with all parameters
 ``corrected'' to $z=0.1$.  That is, using each galaxy's redshift, its
 colors were $K$-corrected to the rest frame, but to $ugriz$ bandpasses
 {\em shifted blueward by a factor $(1+0.1)$ in $\lambda$}.  The absolute
 magnitude and surface brightness are also for the $(z=0.1)$-shifted
 $i$-band.  \citet{blanton03a} produced arrays (e.g. their Fig. 7) of
 two-dimensional galaxy distributions for each pair of the seven
 properties listed, and discussed in depth the features of these bivariate
 distributions.  The plots along the diagonal of their Fig. 7 are
 one-dimensional distributions of each property.

 We use sample14 to generate similar plot arrays at a range of redshifts
 (Figs. 1-4), but we choose to use the apparent magnitude $i$,
 $K$-corrected and corrected for cosmological surface brightness dimming,
 instead of a band-shifted $M_i$.  Thus all properties plotted are
 photometric observables for the galaxies in question,
 shifted to a common redshift.  $K$-corrections are performed using the
 IDL code Kcorrect v3\_2 \citep{blanton03b}.  As in \citet{blanton03a},
 all magnitudes are {\em Petrosian} magnitudes
 \citep[see descriptions in][]{blanton01,strauss02}, which measure a
 fraction of the galaxy light that is constant with distance or size
 (ignoring the effect of seeing); \citet{graham05b} have described a simple
 method for converting Petrosian magnitudes to total magnitudes.

 Note that Figs. 1-4, like \citet{blanton03a}'s Fig. 7, attempt to show
 what a true sample of galaxies at the indicated redshift looks like;
 this is achieved by weighting each galaxy by $1/V_{max}$, where
 $V_{max}$ is ``the volume covered by the survey in which this galaxy
 could have been observed'' \citep{blanton03a}.  This weighting accounts
 for the window function of the survey and the redshift distribution of
 the galaxies in the sample; \S3.4 of \citet{blanton03a} provides further
 details.  As a result of this weighting, our 1-D $i$-distributions have
 the form of Schechter functions, but with a sharp drop at the faint end
 due to the absolute magnitude cut described above (the drop-off is not
 vertical because the cut was performed in the $r$-band).

 Comparing Figs. 1-4 reveals the changes in photometric properties that
 occur as the same sample of galaxies is observed at different redshifts.
 These changes are plotted directly in Figs. 5-6.  Five randomly selected
 galaxies that appear faint and blue (at $z=0.1$) and have exponential
 profiles are plotted at a range of redshifts (Fig. 5); the same is done
 separately for five randomly selected bright ($L^*$), red, de Vaucouleurs
 galaxies (Fig. 6).  The plots along the diagonal of each figure have
 redshift $z$ increasing along the horizontal axis.

 By comparing Figs. 5-6 with Fig. 2, one sees that the S\'ersic index is
 a very useful parameter for red galaxy photo-$z$'s,
 since it is constant with redshift while all other properties are not, and
 red galaxies exhibit a wide range in $n$.  That is, the trajectory along
 which a red galaxy moves in redshift (Fig. 6) is roughly perpendicular to
 the galaxy distribution in all the 2-D plots containing $n$.  The $i$-band
 apparent magnitude is also clearly a useful property when combined with any
 of the other observables: it changes strongly with redshift, and the red and
 blue galaxy trajectories never overlap in the 2-D plots.  Note that there are
 degeneracies in some of the color-color plots (i.e., high-$z$ blue galaxies
 look like low-$z$ red galaxies), particularly those incorporating $r$-band
 data but not $u$-band data.  However, the other colors and the apparent
 magnitude clearly are sufficient to break the degeneracy.

\section{THE PHOTO-$Z$ CODE}
\subsection{Theory}
 We can determine a galaxy's redshift by combining its apparent (observable)
 properties with {\em absolute} quantities, i.e. by specifying
 its {\em type} $T$.  Thus, for a given galaxy targeted for photo-$z$
 measurement, we want to find the peak of $P(T)$, the probability
 distribution of galaxy types that it could be.  This information will allow
 us to compute its redshift.

 The starting assumption of our photo-$z$ technique is that the
 (shifted) {\em empirical} galaxy distributions of \S2.2 can be used as
 {\em probability} distributions.  That is, we want to use the 7-D
 distribution of the previously named observables (of which Figs. 1-4 show
 2-D projections),
 corrected to a given redshift $z$, to approximate $P(T|z)$, the probability
 distribution of galaxy types at that redshift.  If the redshift
 corrections are reliable, then this should be a fairly good approximation
 given the large sample size. According to Bayes' Theorem, a photo-$z$ can
 then be computed as the redshift that maximizes
\begin{equation}
 P(T) = P(T|z) * P(z),
\end{equation}
 where $P(z)$ is the total probability distribution of redshifts {\em for
 the sample of target galaxies}.  Estimating this function well will be
 an important step in applying this photo-$z$ method to any new target
 sample.

 The 7-D distributions are generated across a range of redshifts
 that is believed to cover all galaxies in the target sample, with an
 interval between the redshifts that is less than the rms error of the
 photo-$z$'s.  At each redshift, a target galaxy falls somewhere in
 the $P(T|z)$ distribution, and the value $P(T|z) * P(z)$ is computed
 and stored for comparison to values at other redshifts.

 Initially, a slightly different approach was considered: only one
 distribution would be generated, and each target galaxy would be assigned
 many different redshifts in turn.  Roughly speaking, the best-fit
 photo-$z$ would then be that which places the target at the highest point
 in the distribution.  However, Figs. 1-4 demonstrate that the
 distributions change shape with redshift, so information would be lost
 with this approach.  Furthermore, for reasons described by
 \citet{blanton03b}, ``one can observe a galaxy at $z=0.1$ and reliably
 infer what it would look like at $z=0.3$; it is only the reverse process
 that is difficult.''  Since the median redshift of the LSS samples is
 $z\sim0.1$, we are much better off doing $K$-corrections to the sample
 galaxies than to a target galaxy that may have redshift $z\sim0.3$.
 Finally, the multiple-distribution method is more computationally efficient
 because we can generate the requisite distributions just once and store
 them, so that no $K$-corrections need be performed when we run the code
 on a set of targets.  For all of these reasons, the method of generating
 multiple distributions is favored.

\subsection{Implementation}
 We use IDL to implement the algorithm described above.  Distance
 moduli (for shifting the source galaxies) are computed using the
 cosmological parameters $\Omega_m=0.3$, $\Omega_{\Lambda}=0.7$, and
 $H_0=100$ km/s/Mpc \citep[following][]{blanton03a}.  To avoid assigning
 as photo-$z$'s only those discrete redshifts at which the distributions
 are generated, we interpolate quadratically between the maximizing redshift
 and its immediate neighbors at higher and lower $z$.  We assign the $z$-value
 corresponding to the peak of the fit parabola.  For galaxies assigned the
 minimum or maximum redshift tested, we simply use that value; however, the
 redshift range can always be expanded so that there are few of these
 cases.

 The shifted galaxies are placed into cells in a 7-D array, each dimension
 of which spans a range broad enough to include virtually every galaxy in
 the source sample, at every redshift to be tested.  Given this broad range,
 we must have a large number of cells in each dimension in order to have
 reasonably high type-resolution.
 However, the resolution is limited by both the amount of memory available on
 the system on which the code is run (this is a real problem for 7-D arrays of
 numbers that can become fairly large near a peak in the distribution), and
 2) the fact that the number of points (source galaxies) that go in the
 array is fixed, so that increasing resolution makes the array more and more
 sparsely populated.

 We balance these competing factors by using a resolution of 15
 cells per dimension.  However, for a typical distribution generated at
 this resolution (in particular, for $z=0.01$), only $\sim0.03\%$ of the
 cells in the array are populated, and the majority of these contain just
 one galaxy.  Therefore an adaptive mesh is implemented, ``smoothing'' each
 single-galaxy cell across all neighboring cells.  Specifically, the occupation
 number of each cell is multiplied by a (large) constant $N$, and then all
 cells that lie within one unit (in any combination of dimensions) of a
 single-galaxy cell are populated with numbers, the total of which---for any
 given single-galaxy cell---is $N$.  Thus, after the initial multiplication
 by $N$, no points are added to the distribution; it is merely smoothed around
 each cell that formerly contained a single galaxy.

 Furthermore, not all the cells newly populated by this step are given the
 same value, for they lie at different distances in parameter space from the
 central cell (the one that had only one galaxy). For example, a cell that
 has six coordinates in common with the central cell and only one that
 differs by unity is much ``closer'' than a cell with all seven coordinates
 differing by unity from those of the central cell.  Thus we compute the
 center-to-center distance between each cell and the central one (in units
 of a cell), and place values in the cells that are inversely
 proportional to that distance.  The central cell gets the largest value
 of all, though this is greatly reduced from the value it had before
 smoothing.

 After this smoothing is performed, the $z=0.01$ distribution mentioned
 previously populates $\sim3\%$ of the array, an improvement by two orders
 of magnitude.  In the next section, we will see how this change affects
 photo-$z$ measurements.

\section{RESULTS}
 Photometric redshift routines are usually tested by applying them to
 objects with known (i.e., spectroscopic) redshifts.  Since redshifts
 are known for all galaxies in LSS sample14, we can simply trim the
 sample that we use to generate the distributions, and use
 the remaining galaxies as the target sample.  Specifically, we test the
 code on 1/4 of the sample (55,405 galaxies), using only the remaining
 3/4 to generate the distributions.  Distributions are generated over
 the redshift range $0.02 < z < 0.30$, at intervals of 0.02 in $z$ (note
 that the upper limit extends beyond the greatest redshift present in
 our source sample; still, we include $z=0.30$ in order to verify
 that no galaxies are incorrectly assigned such a high redshift).

 As explained in \S3.2, an estimate of $P(z)$ for the target sample is
 needed.  In this special case, $P(z)$ is the same for both the source
 and target distributions.  $P(z)$ is usually ``divided out'' from the
 source population when each galaxy is weighted by $1/V_{max}$.  Instead,
 in this case we can avoid estimating $P(z)$ entirely by giving each
 source galaxy a weight equal to unity, effectively skipping the
 division by $P(z)$.  Then there is no need to multiply by $P(z)$ later,
 for the probability computed from the distribution at each given
 redshift $z$ gives us $P(T)$ directly.  Fig. 7 shows the 2-D 
 projections of a unity-weighted distribution, as used in this particular
 test.
 
 We define $\Delta z \equiv z-z_{phot}$, where $z$ is the spectroscopic
 redshift and $z_{phot}$ is our photo-$z$.  Without the adaptive mesh
 smoothing, this test yields an rms $\Delta z$ of 0.029, with
 systematic offset of essentially zero (mean $\Delta z\sim-0.0005$).  However,
 our {\em failure rate}, i.e. the percentage of galaxies that are not
 assigned a redshift because they do not fall inside an occupied cell at
 {\em any} of the redshifts tested, is $\sim29\%$.  With the
 smoothing incorporated, the failure rate drops to $\sim11.3\%$, which
 should be acceptable for most purposes; the rms $\Delta z$ also improves
 slightly, to $\sim0.0275$.  Fig. 8 is a plot of $z_{phot}$ vs. $z$ for
 all the galaxies here tested.

 In addition, we examine the performance of the photo-$z$ code on red and
 blue galaxies separately, using the ``optimal color separator'' of
 \citet{strateva01}, $u-r=2.22$.  The target sample, thus divided,
 contains 25,296 ``blue'' galaxies and 30,109 ``red'' galaxies.  The rms
 $\Delta z$ for the red galaxies is $\sim0.0246$; for the blue galaxies,
 it is $\sim0.0303$.  Interestingly, the red galaxies have a
 notably higher failure rate ($\sim16.5\%$) than the blue galaxies
 ($\sim5.1\%$).  Figs. 9 \& 10 are plots of $z_{phot}$ vs. $z$ for the
 red and blue galaxy subsets, respectively.  Table 1 divides the target
 sample even further, both by $u-r$ color and by $i$-magnitude, and
 shows the variation of rms $\Delta z$ with these parameters.  The errors
 are smaller for the brighter galaxies of all colors, despite the fact
 that the fainter galaxies are more numerous in both the training set and
 target sample.

 Table 2 compares our photo-$z$ accuracy to that achieved by other methods.
 Our rms $\Delta z$ is lower than that obtained by \citet{csabai03} using
 two template-fitting methods and their own hybrid technique, and comparable
 to the results of \citet{connolly95}'s quadratic-fitting approach and
 the support vector machine method of \citet{wadadekar05}.  The
 template-fitting methods also produce significant systematic offsets 
 (underestimates), while our method does not.
 \citet{csabai03} reported rms $\Delta z$ of 0.029 for red galaxies and
 0.04 for blue galaxies, so our method shows the most pronounced improvement
 in the photo-$z$'s for blue galaxies.  \citet{csabai03} used a smaller
 sample of $\sim35,000$ galaxies, but using smaller training sets does not
 significantly increase the errors from our method (Mandelbaum et al., in
 preparation).

 \citet{padmanabhan05} have achieved rms $\Delta z \sim0.03$ using a
 template-fitting approach, but they used the deeper SDSS LRG sample, so
 their results are not directly comparable to ours.

 As Table 2 shows, smaller rms $\Delta z$ has been obtained using the neural
 network technique of \citet{collister04} and two techniques (ensemble model
 and Gaussian process regression) introduced by \citet{way06}.  Other neural
 network methods have similarly attained rms $\Delta z \sim0.02$
 \citep{vanzella04, ball04, dabrusco07}.  However, our method is arguably more
 transparent than the neural network techniques.  The next section
 discusses additional improvements that could further reduce our errors in
 future implementations.

\section{CONCLUSIONS}
 We have described a new method for determining photometric redshifts of SDSS
 galaxies.  The method is empirical, and uses a large spectroscopic sample
 of SDSS galaxies to infer distributions of galaxy properties at a range of
 redshifts.  The best-fit redshift is determined by comparing these
 distributions to a galaxy for which a photo-$z$ is desired.  The properties
 used are the five-band SDSS photometry, along with surface brightness and
 the S\'ersic index.  This represents one of the first alternatives to
 neural networks for deriving photo-$z$'s from imaging information beyond
 the photoelectric fluxes.

 Our test of the method produces rms $\Delta z=0.025$ for red galaxies
 in the Main sample, and rms $\Delta z=0.030$ for blue galaxies.  These
 variances are an improvement over those achieved by template-fitting and
 hybrid photo-$z$ codes previously applied to SDSS galaxies, but are somewhat
 worse than the errors typical of neural network methods.

 Implementing an adaptive mesh reduces our method's failure rate, but has
 only a small effect on the rms $\Delta z$, so further adjustments to the
 smoothing technique alone would not likely reduce our errors.  Similarly,
 training sets even larger than the 166,212 galaxies used in our test are
 unlikely to improve the errors significantly (Mandelbaum et al., in
 preparation).  Because our errors are
 currently larger than the redshift spacing (0.02 in $z$) used in generating
 the arrays for the test described here, generating the arrays at finer
 intervals does not by itself reduce our errors.

 One modification that may
 help would be to change the cell spacing for various observables in the
 array---e.g., for the S\'ersic index, cells could be evenly spaced in
 log($n$) rather than evenly spaced in $n$.  Alternatively, the spacing could
 be chosen (for any or all observables) such that
 the peaks in the distribution are spread across many cells, effectively
 providing higher resolution in $P(T|z)$.  This approach would have the
 added advantage of populating a larger fraction of the array, potentially
 reducing the failure rate.

 Looking ahead, the next major challenge for photometric redshift techniques
 (including our
 own) is to make them applicable to higher-redshift galaxy samples.   At
 redshifts only a little higher than the maximum for our sample, the
 intrinsic evolution of the target galaxies becomes significant.  
 This evolution can be calculated with some reasonable confidence for the red, 
 passively evolving galaxies, but not for the actively star-forming blue ones.

 In any case, it is clear that to extend the present techniques to higher
 redshifts, evolutionary corrections will have to be applied if one wishes
 to use the SDSS Main sample to generate the 7-dimensional probability
 arrays. Of course, this approach will require one to estimate the redshift
 distribution $P(z)$ of the target sample in order to compute the individual
 galaxy redshifts.  Alternatively, deeper surveys covering the larger
 redshifts could  be used to generate a high-$z$ training set, but the
 necessity to populate the arrays {\it and} determine
 evolutionary effects self-consistently demands very large datasets. It
 is likely that moderate-sized deep surveys can be used to verify 
 empirical evolutionary corrections to the SDSS Main sample for higher-redshift
 photo-$z$ estimates, and this is the path now being pursued here (Mandelbaum
 et al., in preparation).
 There are several redshift surveys deeper
 than the SDSS spectroscopic sample that overlap with SDSS imaging, including
 the DEEP2 survey \citep[e.g.][]{davis05} and the CNOC2 survey
 \citep[e.g.][]{lin98}, which can be used in this endeavor,
 allowing us to probe more deeply the spatial distribution
 of galaxies throughout the second half of cosmic history.

\acknowledgements
 The authors acknowledge Rachel Mandelbaum, Michael Blanton, David Schlegel,
 and Michael Strauss for useful conversations. David Koo, Nikhil Padmanabhan,
 Alister Graham, and an anonymous reviewer provided helpful comments that
 have improved the manuscript. The photo-$z$ technique
 described herein was originally conceived by David Schlegel. JJW acknowledges
 the Fannie \& John Hertz Foundation Fellowship and the NSF Graduate Research
 Fellowship for partial support of this work. JEG acknowledges support from
 the Eugene Higgins Trust.

 Funding for the SDSS and SDSS-II has been provided by the Alfred P. Sloan
 Foundation, the Participating Institutions, the National Science Foundation,
 the U.S. Department of Energy, the National Aeronautics and Space
 Administration, the Japanese Monbukagakusho, the Max Planck Society, and the
 Higher Education Funding Council for England. The SDSS Web Site is
 http://www.sdss.org/.

 The SDSS is managed by the Astrophysical Research Consortium for the
 Participating Institutions. The Participating Institutions are the American
 Museum of Natural History, Astrophysical Institute Potsdam, University of
 Basel, University of Cambridge, Case Western Reserve University, University of
 Chicago, Drexel University, Fermilab, the Institute for Advanced Study, the
 Japan Participation Group, Johns Hopkins University, the Joint Institute for
 Nuclear Astrophysics, the Kavli Institute for Particle Astrophysics and
 Cosmology, the Korean Scientist Group, the Chinese Academy of Sciences
 (LAMOST), Los Alamos National Laboratory, the Max-Planck-Institute for
 Astronomy (MPIA), the Max-Planck-Institute for Astrophysics (MPA), New Mexico
 State University, Ohio State University, University of Pittsburgh, University
 of Portsmouth, Princeton University, the United States Naval Observatory, and
 the University of Washington.

\clearpage
\begin{table*}
\caption{\label{tab1}
 Our Photo-$z$ Errors as a Function of Color and Apparent Magnitude
}
\begin{center}
\begin{tabular}{c|ccccccc} \tableline \tableline
           &        &            &            &   $u-r$    &            &            &         \\
$i$        & $<1.5$ &   1.5-1.75 &   1.75-2.0 &   2.0-2.25 &   2.25-2.5 &   2.5-2.75 & $>2.75$ \\ \tableline
$<15.5$    & 0.0130 &     0.0156 &     0.0184 &     0.0209 &     0.0181 &     0.0171 &  0.0177 \\
15.5-16    & 0.0185 &     0.0212 &     0.0224 &     0.0247 &     0.0220 &     0.0200 &  0.0199 \\
16-16.25   & 0.0226 &     0.0263 &     0.0250 &     0.0272 &     0.0255 &     0.0199 &  0.0216 \\
16.25-16.5 & 0.0256 &     0.0270 &     0.0311 &     0.0298 &     0.0285 &     0.0229 &  0.0225 \\
16.5-16.75 & 0.0272 &     0.0317 &     0.0335 &     0.0319 &     0.0274 &     0.0243 &  0.0239 \\
16.75-17   & 0.0270 &     0.0319 &     0.0341 &     0.0363 &     0.0300 &     0.0251 &  0.0242 \\
$>17$      & 0.0317 &     0.0343 &     0.0363 &     0.0380 &     0.0334 &     0.0266 &  0.0265 \\ \tableline
\end{tabular}
\tablecomments{
 Most (color, magnitude) bins contain between 500 and 2000 galaxies;
 the least populated bin ($u-r<1.5$, $i<15.5$) contains 189 galaxies.
}
\end{center}
\end{table*}

\begin{table*}
\caption{\label{tab2}
 Comparison of Photo-$z$ Errors from Different Techniques
}
\begin{center}
\begin{tabular}{rrrrrrr} \tableline \tableline
Method            & rms $\Delta z$ & Source              \\ \tableline
CWW templates     & 0.067          & \citet{csabai03}    \\
BC templates      & 0.055          & \citet{csabai03}    \\
Hybrid            & 0.035          & \citet{csabai03}    \\
{\em Our Method}  & {\em 0.0275}   & {\em This work}     \\
SVM               & 0.027          & \citet{wadadekar05} \\
Quadratic fitting & 0.026          & \citet{way06}       \\
Gaussian Process  & 0.023          & \citet{way06}       \\
ANNz              & 0.019          & \citet{way06}       \\
Ensemble model    & 0.019          & \citet{way06}       \\ \tableline
\end{tabular}
\tablecomments{
 Photo-$z$ errors of our method compared to those produced by other methods
 on similar large catalogs of SDSS Main sample galaxies.  The first two
 methods used the spectral templates of \citet{coleman80} and
 \citet{bruzual83b}, respectively.  The quadratic fitting method is similar
 to that introduced by \citet{connolly95}.  The ANNz neural network code is
 that presented by \citet{collister04}.
}
\end{center}
\end{table*}

\clearpage

\begin{figure*}
\resizebox{\hsize}{!}{\includegraphics[angle=0]{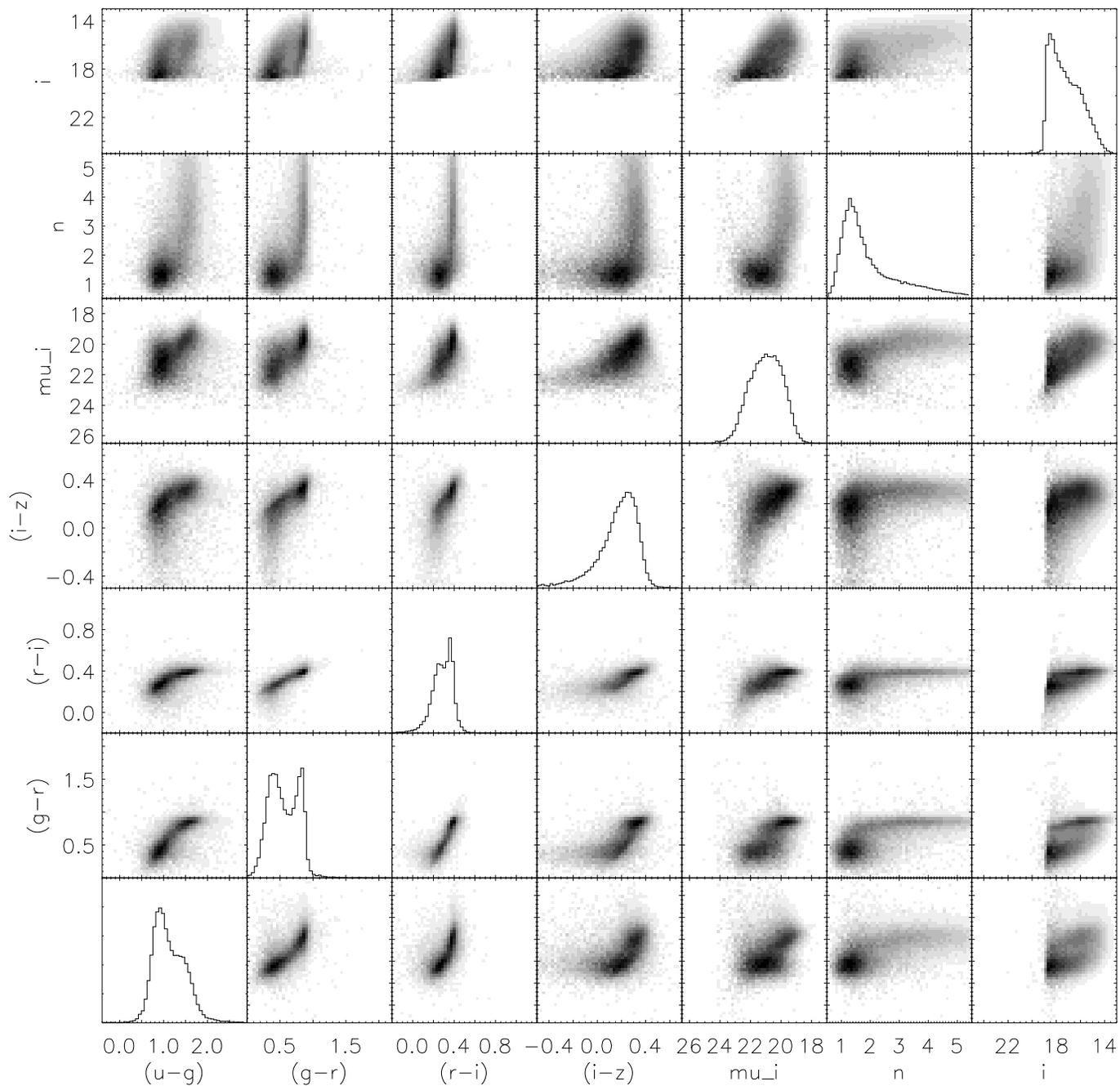}}
\vskip+0.5in
\caption{\label{fig1}
  Properties of sample14 galaxies, $K$-corrected and corrected for
  cosmological surface brightness dimming, to $z=0.05$; $\mu_i$ is the $i$-band surface brightness, $n$ is the S\'ersic index, and $i$ the $i$-band apparent magnitude.  Galaxies
  are weighted by $1/V_{max}$ (explained in the text, \S2.2). Note
  that each 2-D plot is duplicated (reflected about the diagonal).
  The sharp cutoff that appears in the distribution of $i$-magnitudes
  is due to the $r < 17.5$ cut imposed on sample14.
}
\end{figure*}

\begin{figure*}
\resizebox{\hsize}{!}{\includegraphics[angle=0]{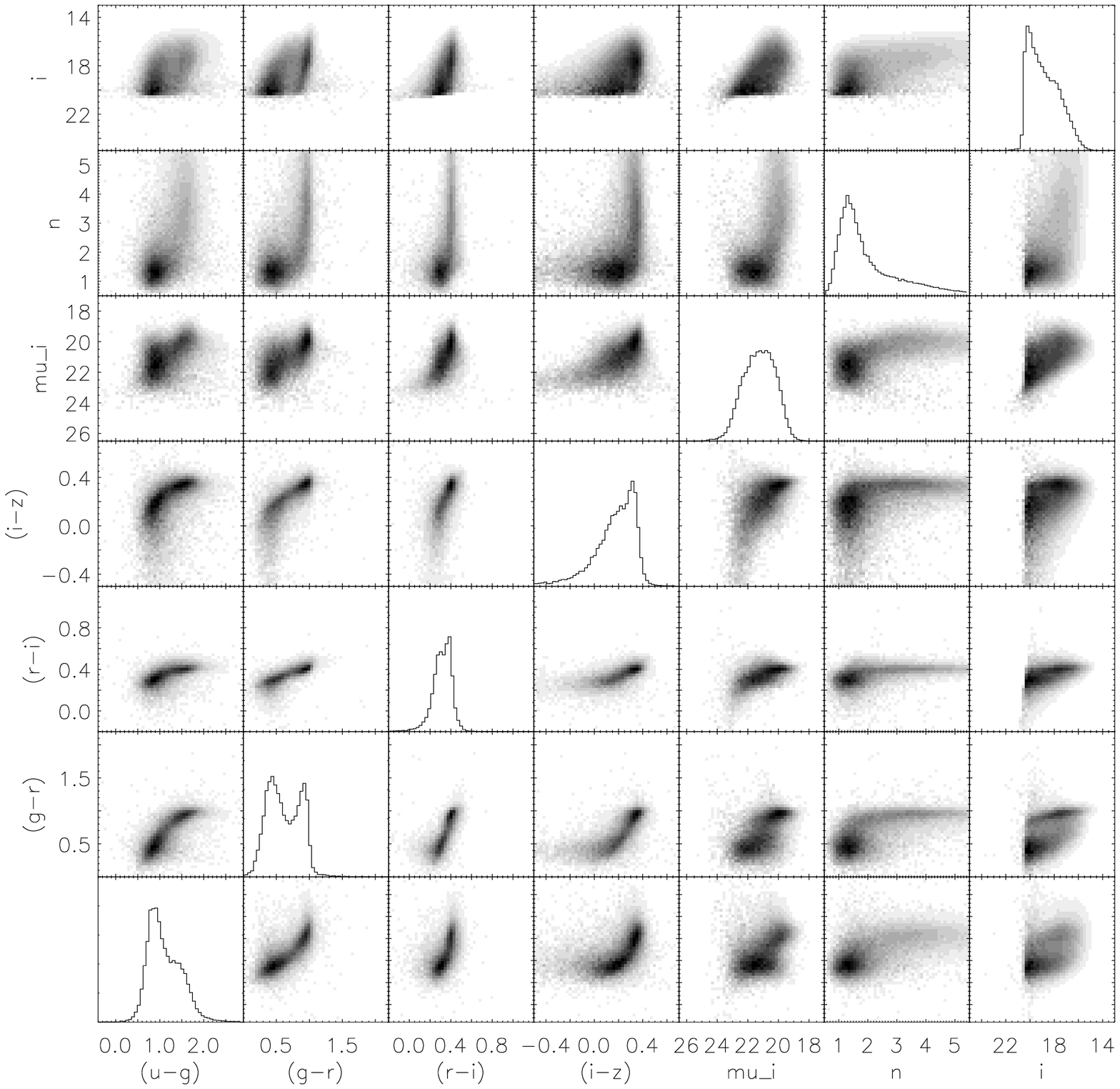}}
\vskip+0.5in
\caption{\label{fig2}
  Same as Fig. 1, but for $z=0.1$.
}
\end{figure*}

\begin{figure*}
\resizebox{\hsize}{!}{\includegraphics[angle=0]{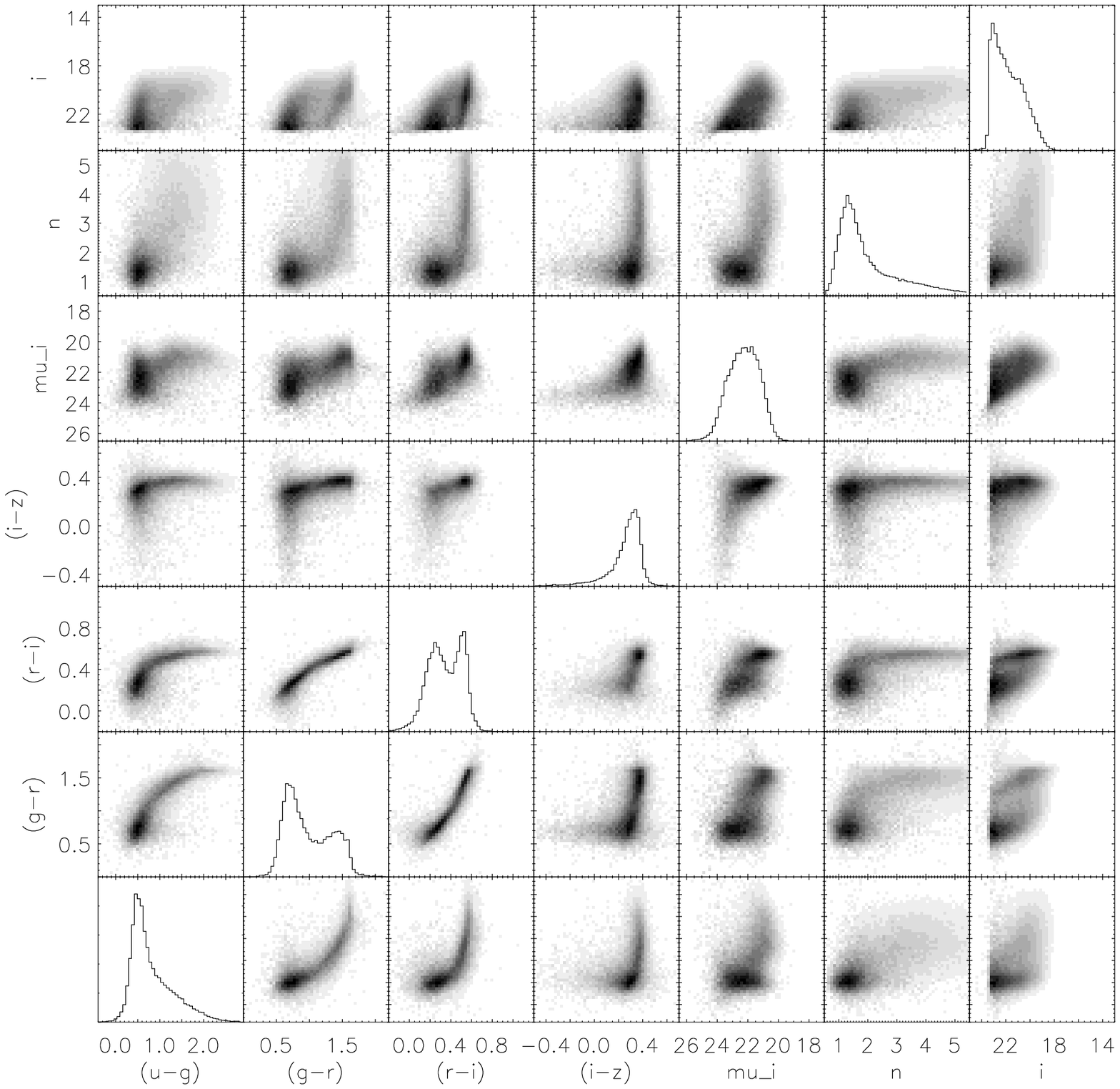}}
\vskip+0.5in
\caption{\label{fig3}
  Same as Fig. 1, but for $z=0.3$.
}
\end{figure*}

\begin{figure*}
\resizebox{\hsize}{!}{\includegraphics[angle=0]{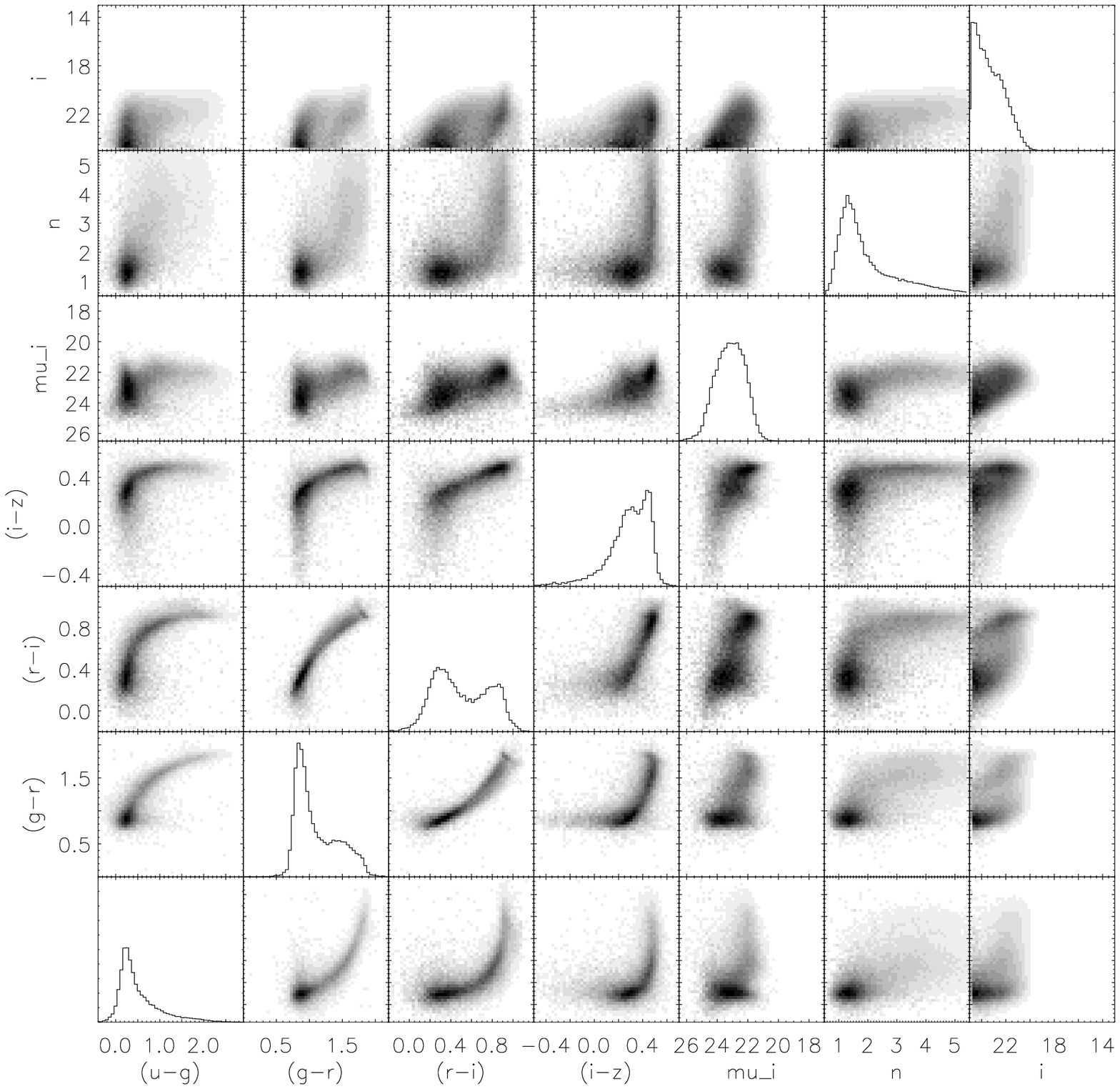}}
\vskip+0.5in
\caption{\label{fig4}
  Same as Fig. 1, but for $z=0.5$.
}
\end{figure*}

\begin{figure*}
\resizebox{\hsize}{!}{\includegraphics[angle=0]{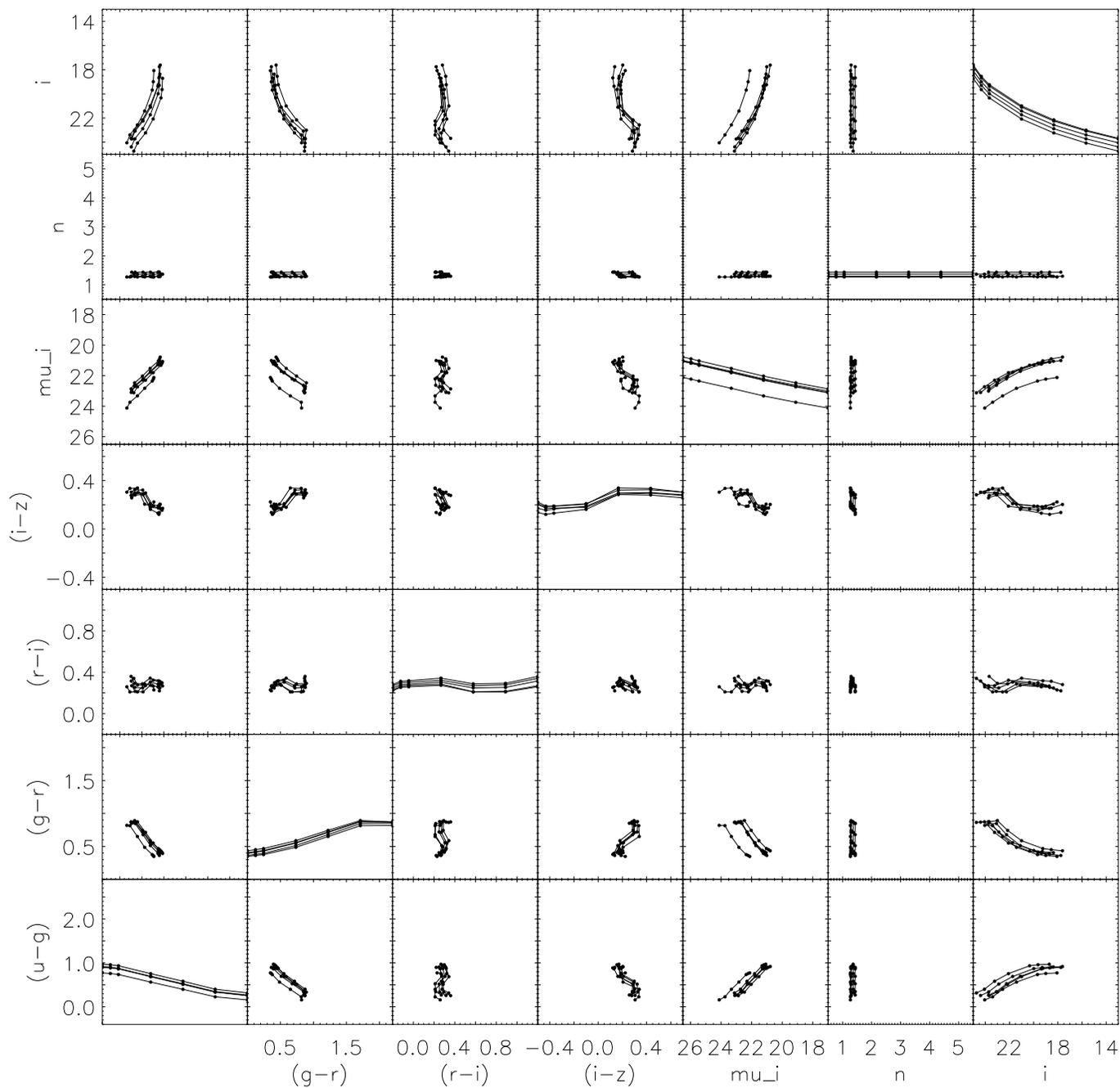}}
\vskip+0.5in
\caption{\label{fig5}
  Seven photometric properties of five randomly selected faint, blue,
  exponential galaxies in the LSS sample14, plotted at a range of
  redshifts (specifically, at $z=0.05, 0.075, 0.1, 0.2, 0.3, 0.4, \&
  0.5$).  The galaxies were selected from a compact 7-D ``box'' at $z=0.1$.
}
\end{figure*}

\begin{figure*}
\resizebox{\hsize}{!}{\includegraphics[angle=0]{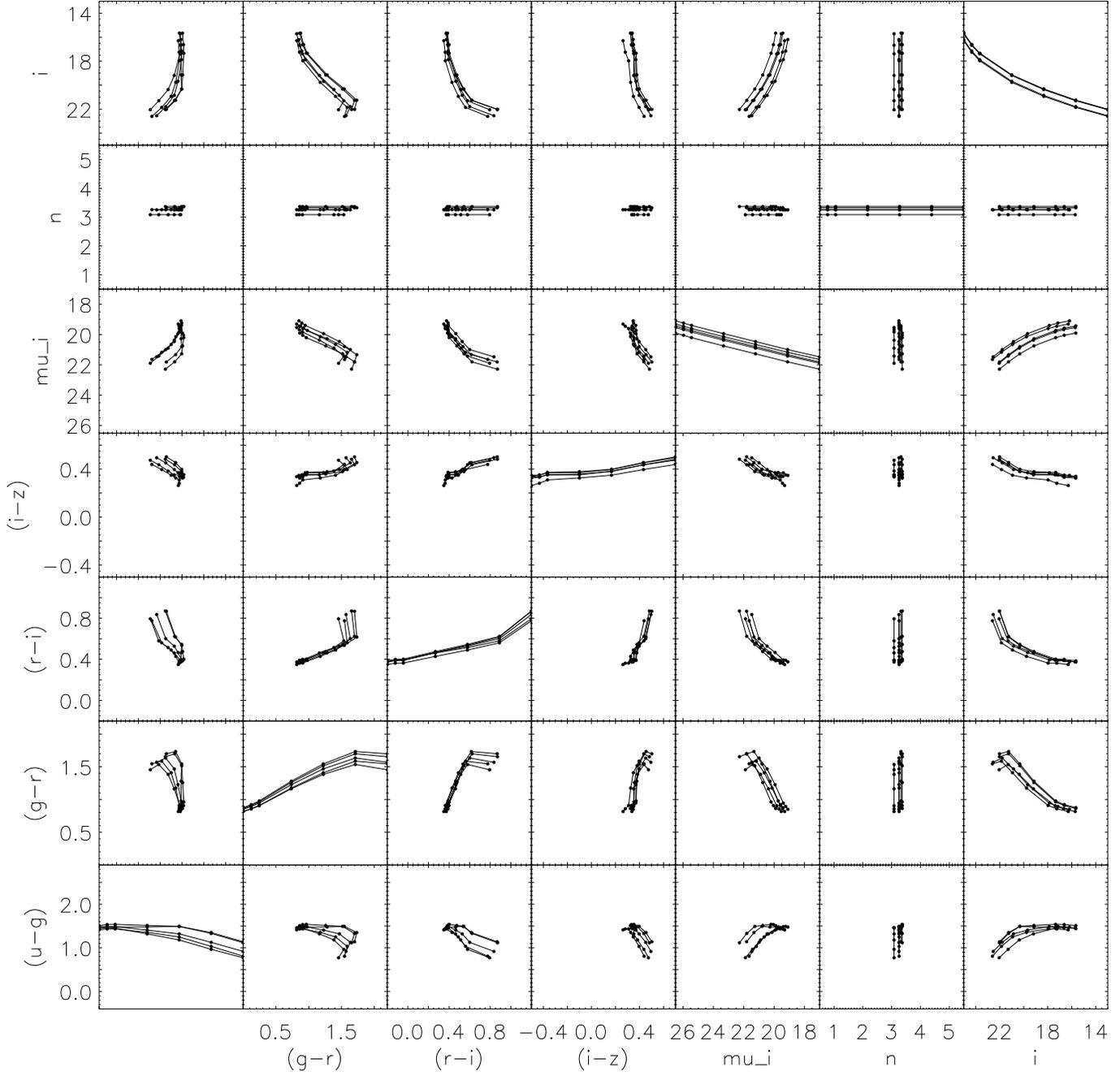}}
\vskip+0.5in
\caption{\label{fig6}
  Same as Fig. 5, but for five randomly selected bright ($L^*$), red, de
  Vaucouleurs galaxies.
}
\end{figure*}

\begin{figure*}
\resizebox{\hsize}{!}{\includegraphics[angle=0]{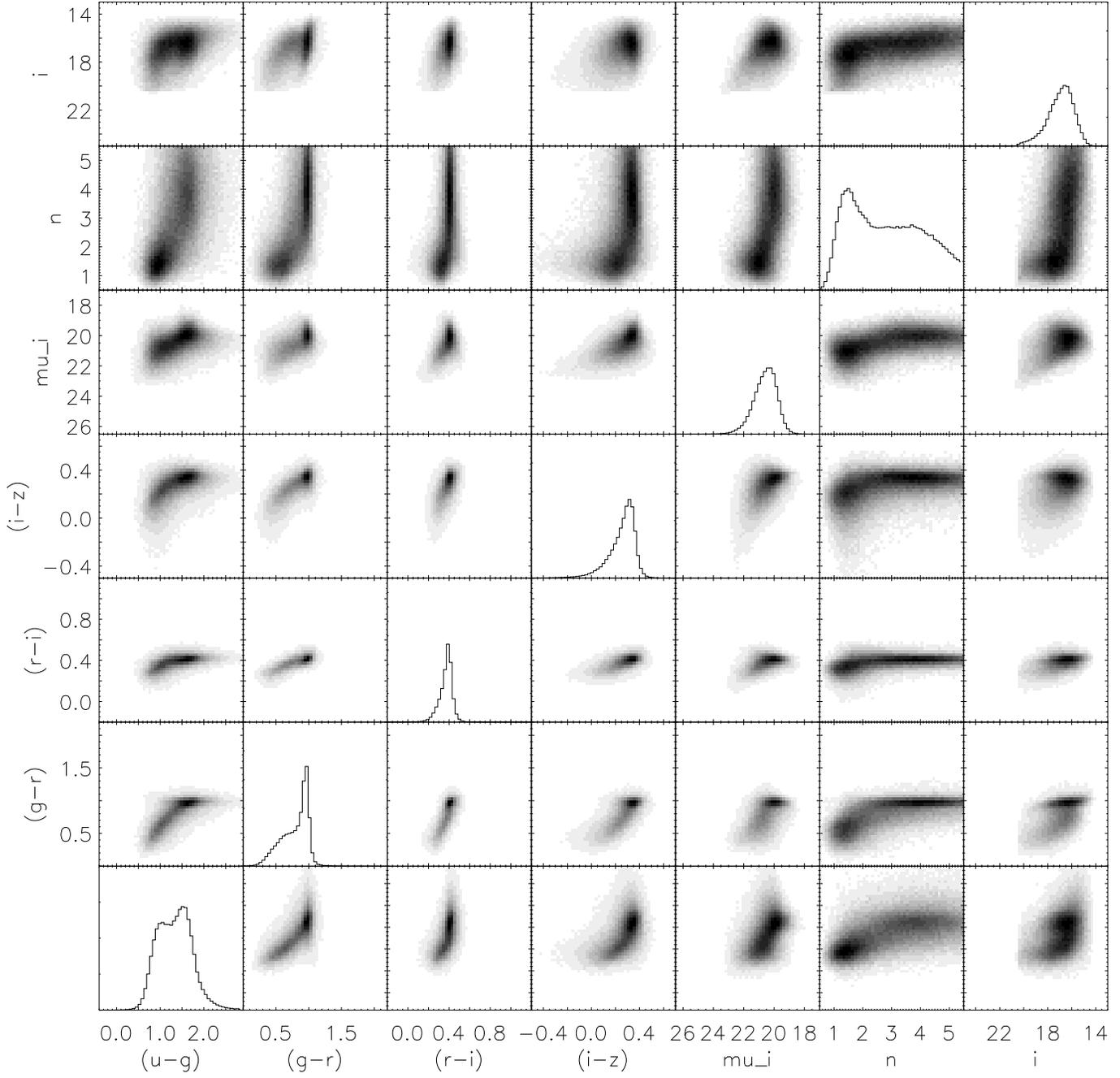}}
\vskip+0.5in
\caption{\label{fig7}
  Same as Fig. 2, but here each galaxy enters the distribution with
  weight 1, instead of $1/V_{max}$.
}
\end{figure*}

\begin{figure*}
\resizebox{\hsize}{!}{\includegraphics[angle=0]{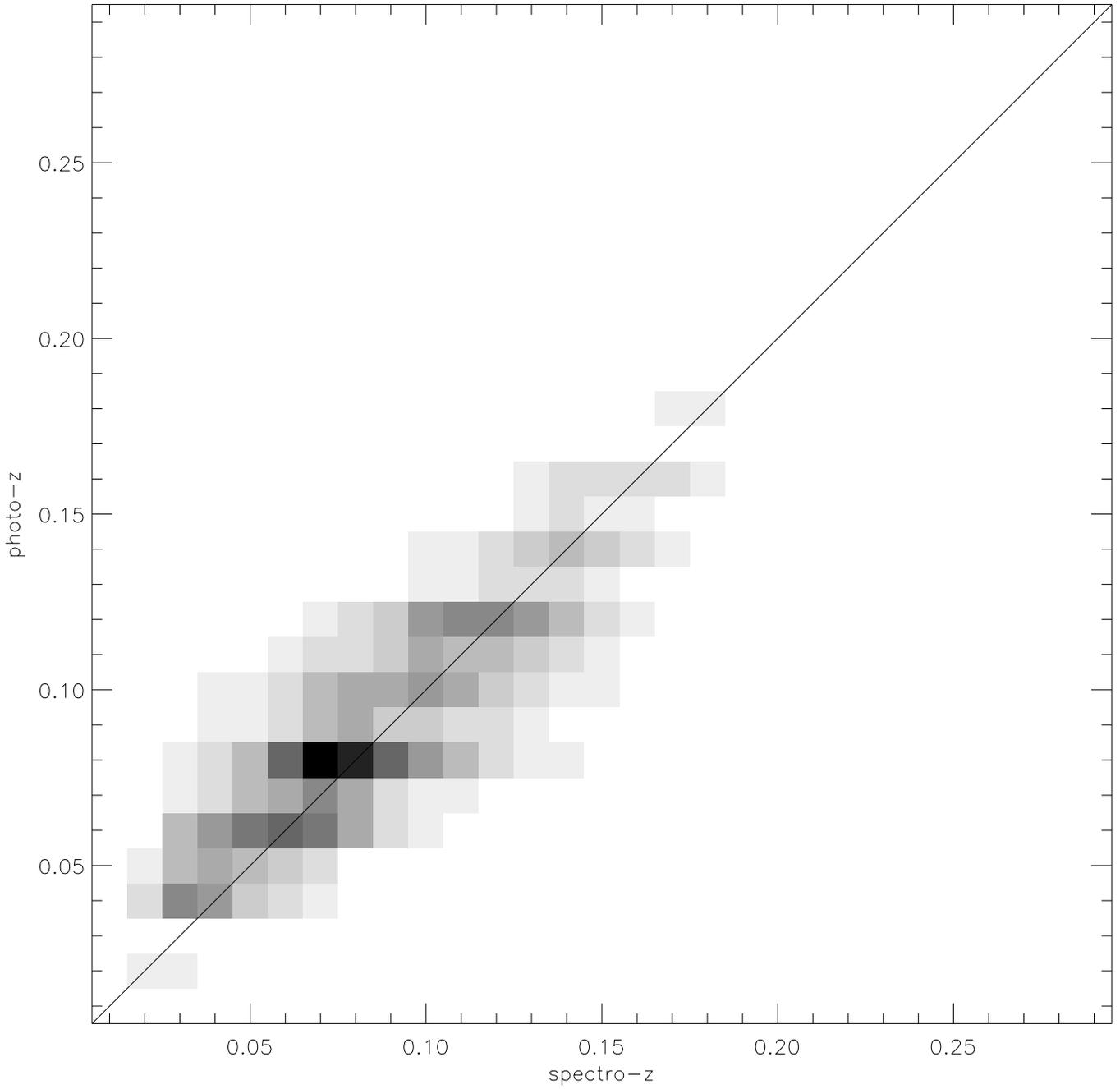}}
\vskip+0.5in
\caption{\label{fig8}
  Our photo-$z$ vs. the spectroscopic $z$ for all galaxies in the
  sample14 subset used for testing, as described in \S4 (49,158
  galaxies).  Rms $\Delta z$ is $\sim0.0275$.
}
\end{figure*}

\begin{figure*}
\resizebox{\hsize}{!}{\includegraphics[angle=0]{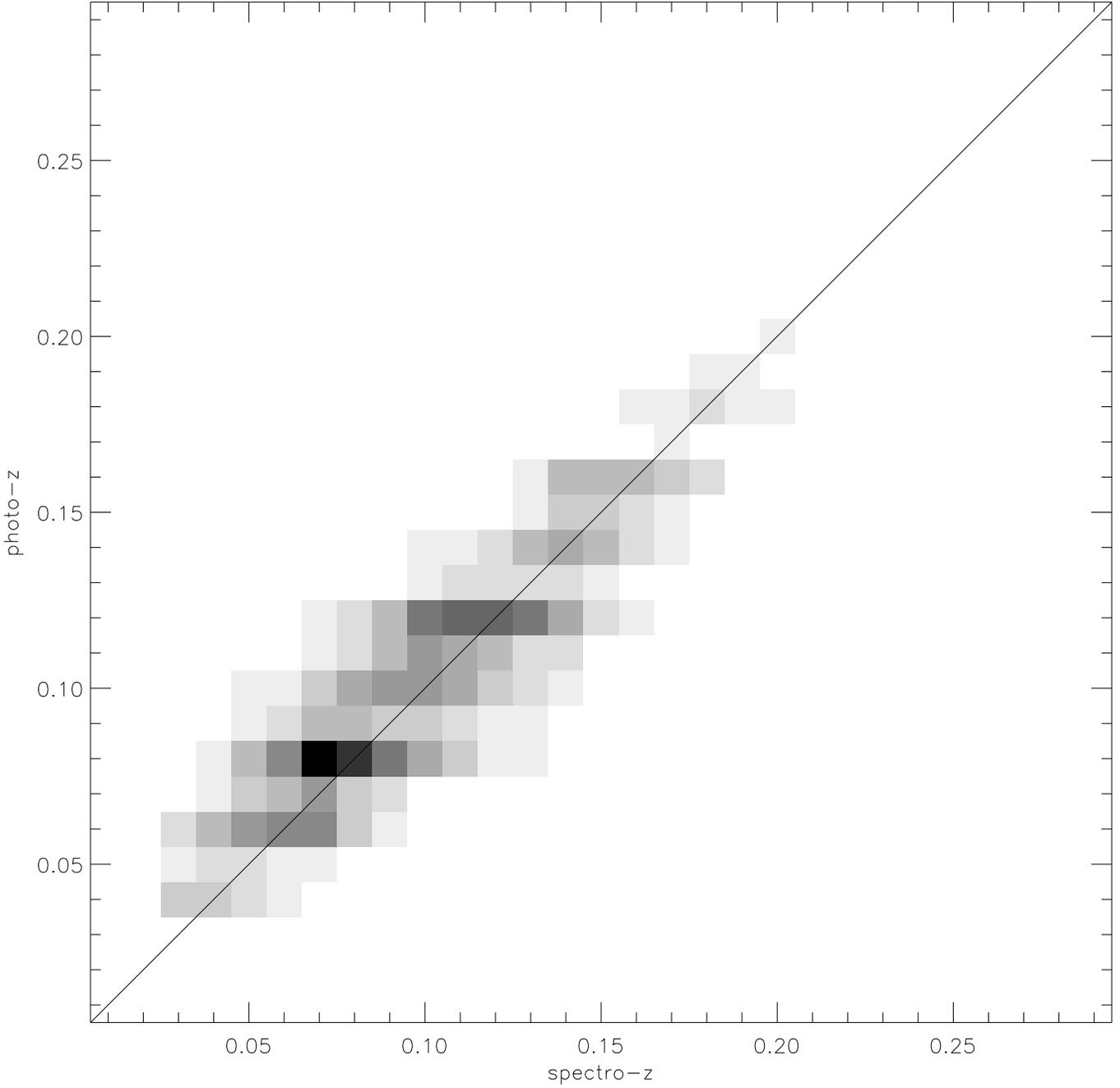}}
\vskip+0.5in
\caption{\label{fig9}
  Same as Fig. 8, but with only the red galaxies (those with $u-r>2.22$)
  plotted (25,146 galaxies).  Rms $\Delta z$ is $\sim0.0246$.
}
\end{figure*}

\begin{figure*}
\resizebox{\hsize}{!}{\includegraphics[angle=0]{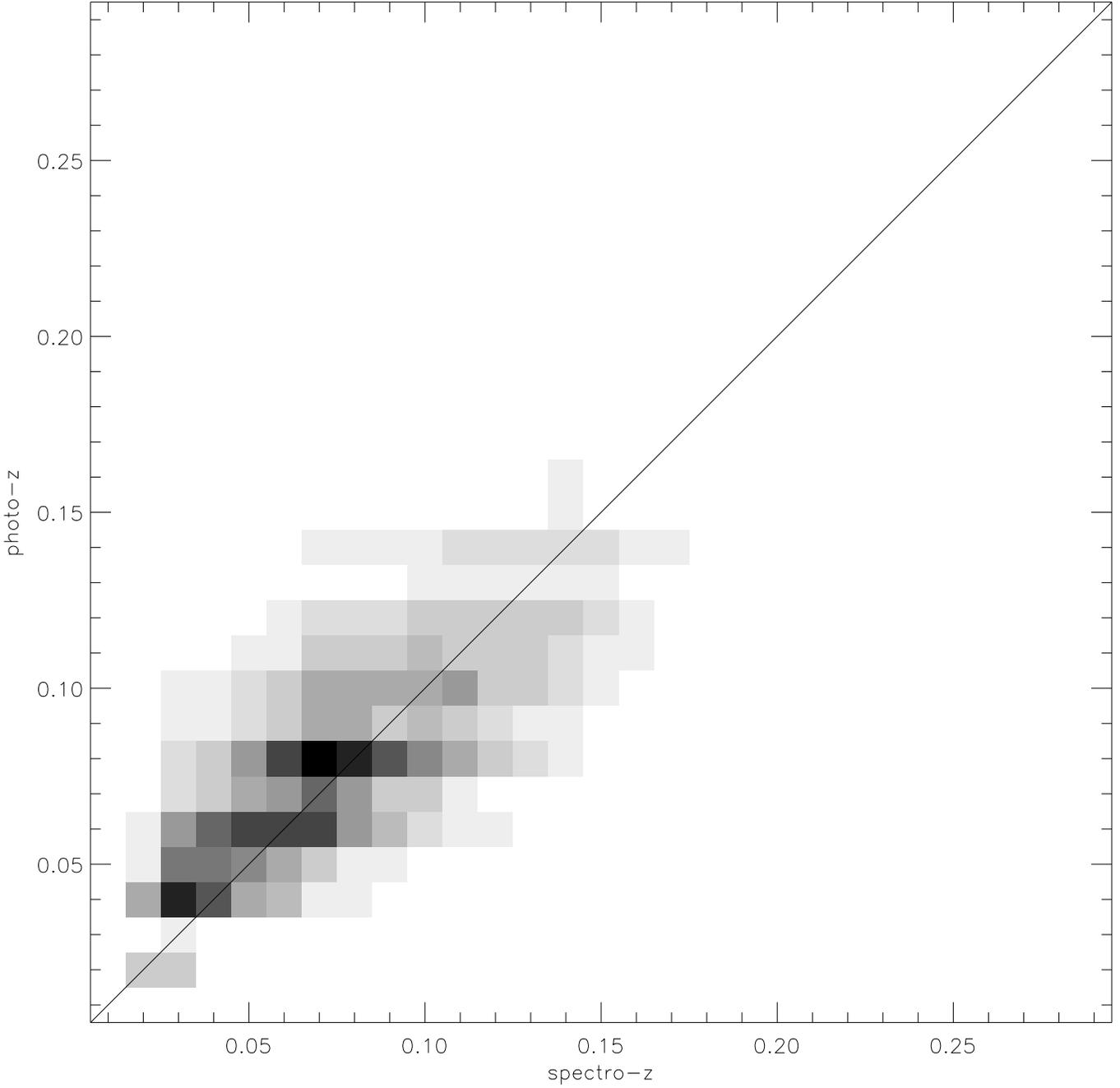}}
\vskip+0.5in
\caption{\label{fig10}
  Same as Fig. 8, but with only the blue galaxies (those with $u-r<2.22$)
  plotted (24,012 galaxies).  Rms $\Delta z$ is $\sim0.0303$.
}
\end{figure*}

\end{document}